\documentclass[11pt,draft]{article}
\usepackage{amssymb,amsmath}
\usepackage{color}
\usepackage{graphicx}
\newcounter{proposition}

\newcommand{\nothing}[1]{}

\newcommand{\beq}[1]{\begin{equation}\label{#1}}
	\newcommand{\eeq}{\end{equation}}
\newcommand{\req}[1]{(\ref{#1})}
\newcommand{\bmu}[1]{\begin{multline}\label{#1}}
	\newcommand{\emu}{\end{multline}}

\newcommand{\eq}{\triangleq}

\renewcommand{\emptyset}{\varnothing}
\newcommand{\str}[1]{\left(\strut{#1}\right)}
\newcommand{\strf}[1]{\left\{\strut{#1}\right\}}

\renewcommand{\varlimsup}{\mathop{\overline{\lim}}\limits}

\newcommand{\x}{{\textbf{\textit{x}}}}

\renewcommand{\a}{{\textbf{\textit{a}}}}

\newcommand{\p}{{\sf p}}

\newcommand{\G}{{\bf G}}
\newcommand{\F}{\mathcal{F}}

\newcommand{\D}{{\cal D}}

\newcommand{\B}{{\bf B}}

\newcommand{\A}{{\cal A}}

\renewcommand{\a}{{\bf a}}

\renewcommand{\r}{{\bf r}}
\renewcommand{\u}{{\bf u}}
\renewcommand{\v}{{\bf v}}
\renewcommand{\S}{{\cal S}}
\renewcommand{\P}{{\cal P}}
\renewcommand{\L}{{\bf L}}
\newcommand{\E}{{\bf E}}

\renewcommand{\l}{\ell}

\renewcommand{\S}{{\mathcal{S}}}
\renewcommand{\L}{{\mathcal{L}}}

\renewcommand{\chi}{\upsilon}
\renewcommand{\l}{{ {L}}}
\newcommand{\0}{{\textbf{\textit{0}}}}
\newcommand{\1}{{\textbf{\textit{1}}}}
\renewcommand{\(}{\left(}
\renewcommand{\)}{\right)}
\renewcommand{\[}{\left[}
\renewcommand{\]}{\right]}

\renewcommand{\l}{\ell}

\begin{document}

	\begin{center}
		{\Large\bf   Almost Cover-Free  Codes and Designs}
		\\[15pt]
		{\bf A. G. D'yachkov, \quad I.V. Vorobyev, \quad N.A. Polyanskii,\quad V.Yu. Shchukin}
		\\[15pt]
		Lomonosov Moscow State University, Faculty of Mechanics and
		Mathematics,\\
		Department of Probability Theory, Moscow, 119992, Russia,\\
		{\sf agd-msu@yandex.ru,\quad vorobyev.i.v@yandex.ru,\quad nikitapolyansky@gmail.com,\quad
			vpike@mail.ru}
	\end{center}
	
	\textbf{Abstract.}\quad An  $s$-subset of codewords of a binary code $X$ is said to be {\em $(s,\ell)$-bad} in $X$  if the  code $X$ contains
	a subset of other  $\ell$ codewords such that the conjunction of the $\ell$ codewords  is covered by  the disjunctive
	sum of the $s$ codewords. Otherwise,  the  $s$-subset of codewords of $X$ is said to  be  {\em $(s,\ell)$-good} in~$X$.
	A binary code $X$  is said to be a cover-free (CF) $(s,\ell)$-code  if the code $X$  does not contain
	$(s,\ell)$-bad subsets. In this paper, we introduce a natural  {\em probabilistic} generalization
	of CF $(s,\ell)$-codes, namely: a binary code is said to be an almost CF $(s,\ell)$-code	if  {\em almost all} $s$-subsets of its codewords are  $(s,\ell)$-good.  We discuss the concept of almost cover-free $(s,\l)$-codes arising in combinatorial group testing problems
	connected with the nonadaptive search of defective supersets (complexes). We develop a random coding method	based on the ensemble of binary constant weight codes to obtain  lower bounds on the capacity of
	such codes. From the main result it is seen that the lower asymptotic bound on the rate for almost CF $(s,\ell)$-codes is essentially greater than the  rate for ordinary CF $(s,\ell)$-codes

	
	\medskip
	
	{\sl Index terms}.\quad {\sf Nonadaptive group testing, search model of defective supersets,
		cover-free codes  and designs,  almost cover-free codes, capacity,
		error probability exponent, random coding bounds}.

\section{Statement of Problem and Results}
		\subsection{Notations and Definitions}
	\quad
	In what follows, the symbol $\eq$ denotes definitional equalities.
	For any positive integer $n$ put $[n]\eq\{1,2,\ldots,n\}$.
	Let $N$ and $t$ be positive integers, $|A|$ -- the size of set $A$.
	The standard symbol  $\lfloor a\rfloor$
	($\lceil a\rceil$) will be used to denote the largest (least) integer $\le a$~($\ge a$).
 Introduce a binary $N\times t$ matrix
	$X=\|x_i(j)\|$ having $N$ rows
	$\x_i\eq\str{x_i(1),x_i(2),\ldots,x_i(t)}$, $i\in[N]$, and $t$
	columns $\x(j)\eq\str{x_1(j),x_2(j),\ldots,x_N(j)}$, $j\in[t]$. 
	  Any such matrix $X$ is called  a {\em binary code of length $N$  and size $t=\lfloor 2^{RN}\rfloor$}
	(briefly, {\em $(N,R)$-code}), where a fixed parameter $R>0$ is called the {\em rate} of
	code~$X$~\cite{g68}.
	A column $\x(j)\in\{0,1\}^N$ is called a {\it $j$-th codeword}.
	The number of $1$'s in column  $x(j)$, i.e.,
	$|\x(j)|\eq\sum\limits_{i=1}^N\,x_i(j)$, is called
	the {\em weight} of  $x(j)$,~$j\in[t]$.
	A code $X$ is called a {\em constant weight} binary code of  weight $w$, $1<w<N$, if for any
	$j\in[t]$, the weight~$|\x(j)|=w$.
	
	For binary vectors $\u\eq(u_1,\ldots,u_N)\in\{0,1\}^N$ and
	$\v\eq(v_1,\ldots,v_N)\in\{0,1\}^N$, we introduce the component-wise
	{\it disjunction} (or {\em disjunctive} (Boolean) sum)
	$\u\bigvee\v$ and {\it conjunction} $\u\bigwedge\v$:
	$$
	\u\bigvee\v\eq\str{u_1\vee v_1,\ldots,u_N\vee v_N},\quad
	\u\bigwedge\v\eq\str{u_1\wedge v_1,\ldots,u_N\wedge v_N},
	$$
	where $0\vee0=0$, $0\vee1=1\vee0=1\vee1=1$,
	$0\wedge0=0\wedge1=1\wedge0=0$, and $1\wedge1=1$.
	We say that $\u$ is {\it covered} by $\v$ ($\v\succeq\u$) if $\u\bigvee\v=\v$.
		
	\subsection{Almost Cover-Free Codes}
	Let $s$ and $\l$ be positive integers such that $s+\l\le t$ and
$\P_s(t)\eq\{\S\,:\,\S\subset[t],\,|\S|=s\}$ is the collection of all  $s$-subsets
	of the set~$[t]$. Note that the size of the collection  $\P_s(t)$ is $|\P_s(t)|={t \choose s}$.
	
	\textbf{Definition 1.}\quad Let $X=(\x(1),\x(2),\dots,\x(t))$ be an arbitrary binary 
   code of length $N$ and size~$t$.
	A set $\S\in\P_s(t)$ is said to be {\em $(s,\ell)$-bad}
	for the code $X$ if there exists a set $\L$, $\L\subset[t]\setminus\S$ of size $|\L|=\l$
	such that 
		\beq{Sbad}
	\bigvee_{j\in \S}\x(j)\,\succeq\,\bigwedge\limits_{j\in \L}\x(j).
	\eeq
	Otherwise, the set $\S\in \P_s(t)$ is called  an {\em  $(s,\ell)$-good } set for the code~$X$.
In other words, a set $\S$, $\S\in \P_s(t)$,  is   $(s,\ell)$-good 	for the code~$X$
if for any set $\L$, $\L\subset[t]\setminus\S$ of size $|\L|=\l$,
 the conjunction $\bigwedge\limits_{j\in\L}\x(j)$ is not covered by the disjunction~$\bigvee\limits_{k\in\S}\x(k)$.

Let the symbol $\B(s,\l,X)$ ($\G(s,\l,X)$) denote the collection of all $(s,\ell)$-bad ($(s,\ell)$-good) sets 
$\S$,  $\S\in\P_s(t)$ for the code $X$ and $|\B(s,\l,X)|$ ($|\G(s,\l,X)|$) is the size of the corresponding
collection. Obviously, 
$$
0\le|\B(s,\l,X)|\le{t\choose s},\quad 0\le|\G(s,\l,X)|\le{t\choose s},\quad
|\B(s,\l,X)|+|\G(s,\l,X)|={t\choose s}.
$$

	Note an evident statement.
	
\textbf{Proposition 1.}	\textit{For $s\ge2$ and $\l\ge1$, any $(s,\l+1)$-good ($(s,\l)$-bad) set for a code $X$ is $(s,\l)$-good ($(s,\l+1)$-bad) set for the code $X$, i.e., the injections are true: $\B(s,\l,X)\subset\B(s,\l+1,X)$ and $\G(s,\l+1,X)\subset\G(s,\l,X)$.}
	
	\textbf{Definition 2.}\quad Let $\epsilon$, $0\le \epsilon\le1$, be a fixed parameter.
	A code $X$ is said to be an {\em almost cover-free  $(s,\l)$-code} of  
{\em error probability $\epsilon$} or, briefly, {\em CF $(s,\l,\epsilon)$-code}  if
	\beq{B_eps}
	\frac{|\B(s,\l,X)|}{{t \choose s}}\,\le\,\epsilon\quad \Longleftrightarrow\quad
    |\G(s,\l,X)|\ge\,(1-\epsilon)\,{t \choose s}.
	\eeq
	
		\textbf{Example 1.}\quad Consider $5\times 5$ code $X$:
	\beq{example}
		X=
		\begin{array}{|ccccc|}
		1 & 0 & 0 & 1 & 1\\
		0 & 1 & 1 & 1 & 0\\
		0 & 1 & 1 & 0 & 1\\
		0 & 1 & 0 & 1 & 1\\
		0 & 0 & 1 & 1 & 1
		\end{array}
	\eeq
		Then $\G(2,2,X)=\left\{\{1;2\},\,\{1;3\},\,\{1;4\},\,\{1;5\},\,\{2;3\}\right\}$ and $X$ is an CF $(2,2,\frac{1}{2})$-code.
		
	From Definition 2 and Proposition 1, it follows
	
\textbf{Proposition 2.}	\textit{Any CF $(s,\l+1,\epsilon)$-code  is an CF $(s,\l,\epsilon)$-code.}
	
	Actually, we have the similar property of monotonicity for CF $(s,\l,\epsilon)$-codes for the case when the parameter $\l$ is fixed.
	  
\textbf{Proposition 3.}	\textit{Let $s\ge2$ and $\l\ge1$. If $X$ is an arbitrary CF $(s,\l,\epsilon)$-code of size $t$ and length $N$, then there exists an CF $(s-1,\l,\epsilon)$-code $X'$ of size $t-1$ and length $N$.}

\textbf{Proof of Proposition 3.}\quad
Let $\B(s,\l,X,i)\eq\left\{\,\S\,:\,i\in S \,\in\,\B(s,\l,X)\right\}$ denote the collection of all $(s,\l)$-bad sets $\S$ for the code $X$, containing the element $i\in[t]$. 
Note that the cardinalities  $|\B(s,\l,X,i)|$,  $0\le|\B(s,\l,X,i)|\le{t-1 \choose s-1}$,   $i\in[t]$,  satisfy the equality:
$$
\sum_{i=1}^t |\B(s,\l,X,i)|=s\cdot|\B(s,\l,X)|\,\le\,s\,{t \choose s}\,\epsilon,
$$
where the last inequality follows from~\req{B_eps}.
This means that there exists $j\in[t]$, such that
$$
|\B(s,\l,X,j)|\le\,\frac{1}{t}\, \,s\,{t \choose s}\,\epsilon\,=\,{t-1 \choose s-1}\,\epsilon.
$$
Then one can  check that the code $X'$ obtained from $X$ by deleting the column  $\x(j)$  is an CF $(s-1,\l,\epsilon)$-code of size $t-1$ and length $N$. 	$\qquad \square$
	
	For the particular case~$\epsilon=0$, 
	the  concept of CF $(s,\l,\epsilon)$-code can be considered as a
	natural probabilistic generalization of the combinatorial concept of 
    {\em cover-free $(s,\l)$-code}
    that is defined in~\cite{e82}-\cite{d02} as the {\em incidence matrix} of a {\em family of finite sets
    in which no intersection of $\l$ sets is covered by the union of $s$ others}.  For the case $\l=1$, cover-free  codes and their 
applications were introduced in~\cite{ks64}.
    For $\l\ge2$, cover-free $(s,\l)$-codes along with their applications
	to {\em key distribution patterns}  were firstly  suggested in~\cite{mp88}.
	      
	Let  $t(N,s,\l)$ be the maximal size of cover-free $(s,\l)$-codes of length $N$ and
	let $N(t,s,\l)$ be the minimal length of cover-free $(s,\l)$-codes of size~$t$.
	Then the number
	\beq{R0}
	R(s,\l)\eq \varlimsup_{N\to\infty}\frac{\log_2 t(N,s,\l)}{N}\,=\,
	\varlimsup_{t\to\infty}\frac{\log_2 t}{N(t,s,\l)}
	\eeq
	is called~\cite{d02} the rate of cover-free $(s,\l)$-codes.
	In the recent papers~\cite{d14_pit,d14_isit},
	one can find a detailed survey of the best known lower and upper bounds on the rate~$R(s,\l)$.
	
	Using the conventional information-theoretic terminology accepted in the probabilistic coding
theory~\cite{g68}-\cite{ck85}, introduce	
	
	\textbf{Definition 3.}\quad Let $R,\,R>0,$ be a fixed parameter. Taking into account 
inequality~\req{B_eps} define  the \textit{error} for  CF $(s,\l,\epsilon)$-codes: 
	\beq{e}
	\epsilon(s,\l,R,N)\eq\min\limits_{X\,:\,t=\left\lceil2^{RN}\right\rceil}\,\left\{\frac{|\B(s,\l,X)|}{{t \choose s}}\right\},\quad R>0,
	\eeq
	where the minimum is taken over all $(N,R)$-codes $X$.  The function
	\beq{E}
	\E(s,\l,R)\,\eq\,\varlimsup_{N\to\infty}\,\frac{-\log_2\epsilon(s,\l,R,N)}{N},\quad R>0,
	\eeq
	is said to be the error  {\em  exponent} for  CF $(s,\l,\epsilon)$-codes, the number   
\beq{Csld}
	C(s,\l)\eq\sup\{R:\,\E(s,\l,R)>0\}
\eeq
	is said to be the \textit{capacity} for  almost CF $(s,\l)$-codes and    
	rate $R(s,\l)$ defined by~\req{R0} is said to be \textit{zero-error} capacity for almost CF $(s,\l)$-codes.
	
For the particular case $\l=1$, Definitions~1-3 were
suggested in our paper~\cite{d14_acct},
in which  we introduce the concept of almost disjunctive  list-decoding codes. The best presently known constructions of such codes were proposed in~\cite{d00_2}. Bounds on the rate for these constructions were computed in the recent paper~\cite{br13}.

Definitions 1-3 and Proposition 1-3 lead to

\textbf{Theorem 1.}  (Monotonicity properties.)  \quad 
{\em The following inequalities hold true}
\beq{mon}
R(s+1,\l)\le R(s,\l)\le R(s,\l-1),\quad C(s+1,\l)\le C(s,\l)\le C(s,\l-1),
$$
$$
\E(s+1,\l,R)\le\E(s,\l,R)\le\E(s,\l-1,R) \quad s\ge1,\quad \l\ge2,\quad R>0.
\eeq

\subsection{Almost Cover-Free  Designs}
\quad	
	By $\hat{\P}_s(\l,t)$ denote the collection of {\it supersets} $\p$, $\p\eq(P_1,P_2,\dots,P_s)$, $P_i\subset \P_\l(t)$, $i\in[s]$,
	where each $\p$ consists of
	$s$ disjoint sets $P\subset [t]$ of size $|P|=\l$, i.e.:
	\begin{equation}\label{P1}
		\hat{\P}_s(\l,t)\eq\strf{\p=(P_1, P_2,\ldots,P_s)\,,
			\begin{array}{c} P_i\subset[t],\,\,|P_i|= \l,\\[3pt]
				P_i\cap P_{j}=\varnothing\mbox{ for $i\ne j$, $i,j\in[s]$, }\end{array}}.
	\end{equation}
	Obviously, the collection $\hat{\P}_s(\l,t)$ 
 has the cardinality
	\beq{car}
	|\hat{\P}_s(\l,t)|= \frac{1}{s!}{t \choose s\l}{s\l \choose (s-1)\l}\cdots{2\l \choose \l}. 
	\eeq
For a superset $\p\in\hat{\P}_s(\l,t)$ and a code $X$,
introduce the binary vector $\r(\p,X)\in\{0,1\}^N$ 
as follows:
	\begin{equation}\label{rpX}
		\r(\p,X)\eq\bigvee_{P\in\p}\bigwedge_{j\in P}\x(j),\quad \r(\p,X)\eq\str{r_1,r_2,\ldots,r_N}.
	\end{equation}
One can see that
	the $i$-th component of $\r(\p,X)$ can be written in the form:
	\begin{equation}\label{r}
		r_i=\begin{cases}
			1, & \text{if there exists $P\in\p$
				such that $x_i(j)=1$ for all $j\in P$},\cr
			0, & \text{otherwise}.\cr
		\end{cases}
	\end{equation}		
	
	\textbf{Definition 4.}\quad Let $X=(\x(1),\x(2),\dots,\x(t))$ be an arbitrary binary
   code of length $N$ and size~$t$. A superset $\p$, $\p\in\hat{P}_s(\l,t)$, is said to be an {\em $(s,\l)$-bad} 
superset for the code $X$,
	if there exists another superset  $\p'\in{\hat{P}_s(\l,t)}$, $\p\neq\p'$,
	such that $	\r(\p,X)=\r(\p',X)$.
	Otherwise, the superset $\p$ is said to be {\em $(s,\l)$-good} superset for the code~$X$.

Let the symbol $\hat{\B}(s,\l,X)$ ($\hat{\G}(s,\l,X)$) denote the collection of all $(s,\ell)$-bad ($(s,\ell)$-good) supersets
$\p$, $\p\in\hat{P}_s(\l,t)$, for the code $X$ and $|\hat{\B}(s,\l,X)|$ ($|\hat{\G}(s,\l,X)|$) is the size of the corresponding
collection. Obviously,
$$
0\le|\hat{\B}(s,\l,X)|\le|\hat{\P}_s(\l,t)|,\; 0\le|\G(s,\l,X)|\le|\hat{\P}_s(\l,t)|,\;
|\hat{\B}(s,\l,X)|+|\hat{\G}(s,\l,X)|=|\hat{\P}_s(\l,t)|.
$$

	\textbf{Definition 5.}\quad Let $\epsilon$, $0\le \epsilon\le1$, be a fixed parameter.
	A code $X$ is said to be an  {\em almost cover-free  $(s,\l)$-design} of \textit{error probability $\epsilon$} or, briefly, CF $(s,\l,\epsilon)$-design	if
	\beq{B'_eps}
	\frac{|\hat{\B}(s,\l,X)|}{|\hat{\P}_s(\l,t)|}\,\le\,\epsilon\quad \Longleftrightarrow\quad
    |\hat{\G}(s,\l,X)|\ge\,(1-\epsilon)\,|\hat{\P}_s(\l,t)|.
	\eeq	

\textbf{Example 2.} For the code $X$ described in~\req{example}, the collection of $(2,2)$-bad supersets
$$
\hat\B(s,\l,X)=\{(\{1;2\},\{4;5\}),\,(\{1;3\},\{4;5\}),\,(\{1;4\},\{2;3\}),\,(\{1;5\},\{2;3\})\}.
$$
It follows that $X$ is an CF $(2,2,\frac{4}{15})$-design.

		\textbf{Definition 6.}\quad Let $R,\,R>0,$ be a fixed parameter. Taking into account
inequality~\req{B'_eps} define  the \textit{error} for  CF $(s,\l,\epsilon)$-designs: 
	\beq{ed}
	\hat{\epsilon}(s,\l,R,N)\eq\min\limits_{X\,:\,t=\left\lceil2^{RN}\right\rceil}\,\left\{\frac{|\hat{\B}(s,\l,X)|}{|\hat{\P}_s(\l,t)|}\right\},\quad R>0,
	\eeq
	where the minimum is taken over all $(N,R)$-codes $X$.  The function
	\beq{E}
	\hat{\E}(s,\l,R)\,\eq\,\varlimsup_{N\to\infty}\,\frac{-\log_2\hat{\epsilon}(s,\l,R,N)}{N},\quad R>0,
	\eeq
	is said to be the error  {\em  exponent} for  CF $(s,\l,\epsilon)$-designs, the number   
	$$
	\hat{C}(s,\l)\eq\sup\{R:\,\hat\E(s,\l,R)>0\}
	$$
	is said to be the \textit{capacity} for almost CF $(s,\l)$-designs.

\medskip	
For the particular case $\l=1$, Definitions 4-6 were already introduced in~\cite{m78} to describe the model called {\em planning screening experiments}.
In~\cite{m78}, it was proved that the capacity of almost CF $(s,1)$-designs $\hat{C}(s,1)=1/s$. One can see that Definitions 4-6 represent a natural generalization of almost CF $(s,1)$-designs. We conjecture that for any $s\ge2$ and $\l\ge2$, the capacity  $\hat{C}(s,\l)=1/(s\,\l)$. 

In Section \ref{PT2}, we establish 

\textbf{Theorem 2.}  (Upper Bounds on Capacities $C(s,\l)$ and  $\hat{C}(s,\l)$)  \quad 
{\em The following inequalities hold true}
\beq{up}
C(s,\l)\le \hat{C}(s,\l)\le 1/(s\,\l), \quad  {\E}(s,\l,R)\,\le\,\hat\E(s,\l,R) 
\quad s\ge1,\quad \l\ge1,\quad R>0.
\eeq

\medskip
However, in spite of the greater capacity, {\em using}  of CF $(s,\l,\epsilon)$-designs for the superset identification problem $\p\in\hat{P}_s(\l,t)$ 
is \textit{practically unacceptable}, since it  requires much greater complexity, which is evidently equal to the complexity of exhaustive search ${|\hat{\P}_s(\l,t)|}\sim t^{s\,\l}$.
It will be shown in Section~\ref{bool} that CF $(s,\l,\epsilon)$-codes are efficient CF $(s,\l,\epsilon)$-designs and for such codes the algorithm of identification supersets $\p\in\hat{\P}_s(\l,t)$,
is essentially faster than the trivial one, and its complexity is proportional to~$t^\l$.

\subsection{Lower Bounds on $R(s,\ell)$, $C(s,\l)$}
	\quad
	The best presently known  upper and lower bounds on the rate  $R(s,\l)$ of cover-free $(s,\l)$-codes were
	presented in~\cite{d14_pit,d14_isit}.
	If $\l\ge1$ is fixed and $s\to\infty$, then these bounds have the following asymptotic  form:
	\beq{upR1-as}
	R(s,\l)\,\le\,
	\frac{(\l+1)^{\l+1}}{2e^{\l-1}}\,
	\frac{\log_2s}{s^{\l+1}}\,(1+o(1)), 
	\eeq
	\beq{lowR1-as}
	R(s,\l)\,\ge\,\frac{(\l+1)^{\l+1}}{e^{\l+1}}\,
	\frac{\log_2s}{s^{\l+1}}\,(1+o(1)).
	\eeq

	In the present paper, we suggest a modification of the random coding method developed
	in~\cite{d14_pit} and~\cite{d14_acct}, which permits us to obtain a lower bound on the capacity $C(s,\ell)$.
	Let
	$$
	[x]^+\eq\begin{cases}
	x & \text{if $x\ge0$}, \cr
	0 & \text{if $x<0$},\cr
	\end{cases}
	\quad\text{and}\quad
	h(a)\eq -a\log_2a-(1-a)\log_2(1-a),\; 0<a<1,
	$$
	denote the positive part function and the binary entropy function.
	In  Section \ref{PT3}, we prove

	\textbf{Theorem 3.}\quad
	(Random coding lower bound~$\underline{C}(s,\ell)$).
	{\em The following two claims hold true.}\quad  \\
	{\bf Claim 1.}\quad
	{\em  For $\l\ge2$ the  capacity $C(s,\ell)$ for almost cover-free codes satisfies inequality
		\beq{CSL}
		C(s,\ell)\ge \underline{C}(s,\ell)\eq\frac{1}{\l}\max\limits_{0\le Q\le 1}\D(\l,Q,\hat q),
		\eeq
		where the function $\D(\ell,Q,\hat q)$ is defined in  the  parametric  form
		\beq{DLQ}
		\D(\l, Q, \hat q) \eq (1-Q)\l\,\log_2 z-(1-\hat q)\log_2[1-(1-z)^\l]+
		\eeq
		$$
		+\ell\left(\frac{(1-Q)}{z}(1-z)-\left(\frac{(1-Q)}{z}-\hat q\right)(1-z)^\l\right)\log_2[1-z]+\l h(Q),
		$$
		and parameters $z$ and $\hat q$ are uniquely determined by the following equations
		\beq{QZL}
		Q=\frac{(1-z)(1-(1-z)^\ell)-(1-\hat q)z(1-z)^\ell}{1-(1-z)^\ell},\qquad \hat q=1-(1-Q)^s.
		\eeq}
	{\bf Claim 2.}\quad {\em For a fixed parameter $\ell\ge2$ and $s\to\infty$, the lower asymptotic bound
		on ${C}(s,\ell)$ is:}
	\beq{ranL-4}
	{C}(s,\l)\,\ge\,\frac{\log_2 e}{s^\l}\cdot\frac{\l^{\l-1}}{e^{\l}}\,(1+o(1)).
	\eeq
	
\subsection{Boolean  Model for Nonadaptive Search of Supersets}\label{bool}
	\quad
Denote by $\P_s(\l,t)$ the following collection
	of {\it supersets} $\p$, $\p\eq(P_1,P_2,\dots,P_k)$, $P_i\subset [t]$, $i\in[k]$, $k\le s$,
	where each superset $\p$ is composed of not more than
	$s$ subsets $P\subset [t]$ of size $|P|\le\l$, i.e.:
	\begin{equation}\label{P}
		\P_s(\l,t)\eq\strf{\p=(P_1, P_2,\ldots,P_k)\,\,:\,\, k\le s,
			\begin{array}{c} P_i\subset[t],\,\,|P_i|\le \l,\\[3pt]
				P_i\not\subseteq P_{j}\mbox{ for $i\ne j$, $i,j\in[k]$, }\end{array}}.
	\end{equation}
		For a superset $\p\in\P_s(\l,t)$ and a code $X$, the vector $\r(\p,X)$ is defined in the same way as in~\req{rpX}.

	\textbf{Definition 7.}~\cite{d02}\quad
	A binary $N\times t$ matrix $X$ is called a
	{\it cover-free $(s,\l)$-design} or, briefly, \textit{CF $(s,\l)$-design} if
	for any $\p',\p''\in\P_s(\l,t)$,
	$\p'\ne\p''$, the vector~$\r(\p',X)\ne\r(\p'',X)$.

	\quad
	Let us first remind the well-known application of CF
	$(s,1)$-designs which is called {\it the boolean search model for sets}
	\cite{ks64}. Suppose a set of $t$ samples is given. We
	identify it with the set $[t]$. Assume we know that some of them are
	{\it positive}. The number of positive samples is bounded above
	by the given integer $s$. Our aim is to detect the whole set of positive
	samples which is referred to as {\it positive set} $P\subset[t]$.
	We use {\it group tests}, i.e., take a subset (group)
	$G\subset[t]$ and check whether $G$ contains at least one positive sample
	(i.e., $G\cap P\ne\emptyset$) or not.
	
	In the present paper we consider a generalization of this model which is called
	{\it the boolean search model for supersets}~\cite{d02}.
	Assume that a {\it positive superset}
	$\p\in\P_s(\l,t)$ is fixed instead of positive set.
	Our aim is to detect it using a number of group tests, where {\em each test
		checks whether a testing group $G$ contains at least one set
		$P\in\p$ or not}.
	One can see that for $\l=1$ each set $P\in\p$ is composed of
	exactly one sample and the model coincides with the boolean
	search model for sets.   
	Now assume that we use $N$ tests. They can be encoded by a
    code~$X=\|x_i(j)|$.
	A column (codeword) $\x(j)$ corresponds to the $j$-th sample;
	a row $\x_i$ corresponds to the $i$-th test. We put $x_i(j)\eq1$ iff
	the $j$-th sample is included into the $i$-th testing group;
	otherwise we put $x_i(j)\eq0$.
	Then it is easy to see that  the {\em outcomes}~\req{r} of all $N$ tests form
	the binary vector $\r(\p,X)$ \req{rpX}, where $\p\in\P_s(\l,t)$ is
	the ({\em unknown}) positive superset. Thus, the code $X$ should be designed
	in such a way that we should be able to detect a superset $\p$ given
	the vector $\r(\p,X)$. Obviously, it is possible if and only if
	$X$ is an CF $(s,\l)$-design  (see Definition~7).
	Note that we  deal with the {\it nonadaptive} search model
	in which we are not allowed to use the outcomes of the previous tests
	to form the future ones. The given boolean search model for supersets
    (also called the search model for complexes) when all tests are performed simultaneously
    arises from the needs of molecular biology. It was firstly suggested in~\cite{t99}.
	
	In addition, one can easily understand the necessity of the additional condition
	in \req{P}: no set $P\subset[t]$ which is an element of a superset
	$\p\in\P_s(\l,t)$, can be included into another set $P'\in\p$. Indeed,
	if this holds, then we can consider another superset $\widehat\p\in\P_s(\l,t)$
	having the form $\widehat\p=\p\backslash\{P'\}$. Evidently, for any binary
	$N\times t$ matrix $X$, the outcomes  $\r(\p,X)$ and
	$\r(\widehat\p,X)$ are identical. Thus, we cannot
	distinguish these supersets. In ~\cite{d02}, we established
	
	\textbf{Proposition 4.}~\cite{d02}\quad
	{\sf1)} Any cover-free $(s,\l)$-code is an cover-free $(s,\l)$-design.
	{\sf2)} Any cover-free $(s,\l)$-design is an cover-free $(s-1,\l)$-code and an cover-free $(s,\l-1)$-code.
	\medskip

	Let $X$ be an arbitrary binary $N\times t$ matrix
	and $\p^{\rm(un)}\in\P_s(\l,t)$ be an {\em unknown} superset. Any fixed set
	$P'\subset[t]$, $|P'|\le\l$,  is called {\it acceptable} for the \textit{known}  vector
	$\r^{\rm(kn)}\eq\r(\p^{\rm(un)},X)$ if
	the conjunction $\bigwedge\limits_{j\in P'}\x(j)$ is covered by~$\r^{\rm(kn)}$, i.e.,
	$$
	\r^{\rm(kn)}\,=\,\r(\p^{\rm(un)},X)\,=\,\bigvee_{P\in\p^{\rm(un)}}\bigwedge_{j\in P}\x(j)\,\succeq\,\bigwedge\limits_{j\in P'}\x(j).
	$$
	An acceptable set $P'$ is called a
	{\em minimal} acceptable set if no subset $P''\subsetneq P'$ is acceptable.
	In the boolean search model for supersets, an effective decoding algorithm
	is based on the following evident
	
	\textbf{Proposition 5.}~\cite{d02}\quad
	If $X$ is an cover-free  $(s,\l)$-code,
	then any superset $\p\in\P_s(\l,t)$ is composed of all
	minimal acceptable sets for the vector $\r(\p,X)$. This means that
	one can uniquely decode $\p^{\rm(un)}$
	on the base of known vector $\r^{\rm(kn)}=\r(\p^{\rm(un)},X)$,
	and the decoding complexity is proportional to
	${t\choose 1}+\cdots+{t\choose\l}$, which does not depend on $s$.
	When $t\to\infty$ and $\l$ is fixed, then this
	complexity $\sim t^{\l}/\l!$.
	\medskip
	
	Note that in the general case of cover-free  $(s,\l)$-design and the trivial
	decoding algorithm,  we need to check  all possible supersets $\p\in\P_s(\l,t)$,
	i.e., calculate the
	vector $\r=\r(\p,X)$ for all possible supersets $\p$ and compare
	this vector with the known result $\r^{\rm(kn)}$. If $s$ and $\l$
	are fixed and $t\to\infty$, then the number of such comparisons
	(decoding complexity) is proportional to
	\beq{sl1}
	|\P_s(\l,t)|\,\ge\, {{t\choose\l}\choose s}
	\sim\frac{t^{s\l}}{s!(\l!)^s}.
	\eeq
	
	Thus, CF  $(s,\l)$-codes form a class of CF   $(s,\l)$-designs for which
	the  decoding algorithm based on Proposition~5 is strongly better than the trivial one.

Let $\l\ge1$ be fixed and $s\to\infty$. Taking into account~\req{lowR1-as} we conclude that for sufficiently large $t$ the {\em use of CF $(s,\l)$-codes} gives
the bounds:
$$
\frac{\log_2s}{s^{\l+1}}\cdot\frac{(\l+1)^{\l+1}}{2e^{\l-1}}\,
\,(1+o(1))\ge\log_2t/N\,\ge\,
\frac{\log_2s}{s^{\l+1}}\cdot\frac{(\l+1)^{\l+1}}{e^{\l+1}}\,(1+o(1)).
$$
In virtue of Theorem 3,  the capacity for CF $(s,\l, \epsilon)$-codes $C(s,\l)$
can be interpreted as the theoretical tightest upper bound on the information rate $\log_2t/N$ with error probability~$\epsilon\to0$. Therefore, the bound \req{ranL-4} means that for $\l \ge 2$, $s\to\infty$ and sufficiently large $t$, {\em using of CF $(s,\l, \epsilon)$-codes} guarantees the inequality:
$$
\log_2t/N\,\ge\,\frac{\log_2 e}{s^\l}\cdot\frac{\l^{\l-1}}{e^{\l}}\,(1+o(1)).
$$

\section{Proof of Theorem 2.}\label{PT2}
	\quad
	For any superset $\p\in \hat{P}_s(\l,t)$, $\p=\{P_1, P_2,\ldots,P_s\}$, define a set $T(\p)$ of its \textit{projections}  $$T(\p)\,\eq\,\big\{\S\in\P_s(t):\quad S=\{a_1,a_2,\dots ,a_s\},\quad a_i\in P_i,\,P_i\in\p,\,i\in[s]\big\}.$$
	 One can see $|T(\p)|=\l^s$. Observe that if all sets $\S\in T(\p)$ are $(s,\l)$-good for the code $X$, then the superset $\p$ is also a $(s,\l)$-good superset for the code $X$.
	
	Assume that a code $X$ is an CF  $(s,\l,\epsilon)$-code. It means that the number \req{B_eps} of bad $(s,\l)$-sets doesn't exceed $\epsilon\cdot{t\choose s}$. Given a bad $(s,\l)$-set $B\in\P_s(t)$  for the code $X$, one can check that the number of $\p\in \hat{P}_s(\l,t)$ such that $B\in T(\p)$ is at most  ${t-s \choose s(\l-1)}{s(\l-1) \choose (s-1)(\l-1)}\cdots{2(\l-1) \choose \l-1}$. This implies that the number of bad $(s,\l)$-supersets is at most $\epsilon\cdot{t \choose s}{t-s \choose s(\l-1)}{s(\l-1) \choose (s-1)(\l-1)}\cdots{2(\l-1) \choose \l-1}$ or $\epsilon\cdot\l^s\cdot|\hat\P_s(\l,t)|$, where $|\hat\P_s(\l,t)|$ is computed in~\req{car}. Therefore, the code $X$ is also an CF $(s,\l,\epsilon\cdot\l^s)$-design. In other words, we proved the relations $C(s,\l)\le\hat C(s,\l)$ and ${\E}(s,\l,R)\,\le\,\hat\E(s,\l,R)$.
	
	\quad Now, fix $R>0$ and $\epsilon>0$ and suppose that the code $X$ is an CF $(s,\l,\epsilon)$-design of
	length~$N$ and size~$t \eq \left\lfloor 2^{RN}\right\rfloor$.
	Observe that for any two various good (see Def.~4) supersets $\p,\,\p'\in\hat{\G}(s,\l,X)$, $\p\neq\p',$ two binary vectors $\r(\p,X)$ and $\r(\p',X)$ defined by~\req{rpX} are distinct, i.e., $\r(\p,X)\neq\r(\p,X)$. Thus, from the definition~\req{B'_eps} of CF $(s,\l,\epsilon)$-design, it follows
	\beq{bound1}
	(1 - \epsilon)\cdot|\hat\P_s(\l,t)|=(1 - \epsilon)\cdot\frac{1}{s!}{t \choose s\l}{s\l \choose (s-1)\l}\cdots{2\l \choose \l} \, \le
	\,  2^N,\quad t = \left\lfloor 2^{RN}\right\rfloor.
	\eeq
	Comparing left and right-hand sides of inequality~\req{bound1} leads to the lower asymptotic bound
	$$
	\hat\epsilon(s,\l, R, N)\ge\, 1-2^N\cdot\left(\frac{1}{s!}{t \choose s\l}{s\l \choose (s-1)\l}\cdots{2\l \choose \l}\right)^{-1}\,=
	\,1-2^{-N[(s\l\cdot R-1)+o(1)]},\quad N\to\infty.
	$$
	This inequality means that the condition $R<1/(s\l)$ is necessary for $\hat\E(s,\l,R)>0$. It follows $ \hat C(s,\l)\le\frac{1}{s\l}$. $\qquad \square$
	
	\section{Proof of Theorem 3}\label{PT3}
\quad
\textbf{Proof of Claim 1.}\quad
For an arbitrary code $X$, the number $|\B(s,\ell,X)|$  of  $(s,\ell)$-bad sets
in the code $X$ can be represented in the form:
\beq{s_L1}
|\B(s,\ell,X)|\eq\sum\limits_{\S\in\P_s(t)}\,\psi(X,\S),\qquad
\psi(X,\S)\,\eq\,
\begin{cases}
	1 & \text{if the set $\S\in\B(s,\l,X)$},\cr
	0 & \text{otherwise}.\cr
\end{cases}
\eeq
Let $Q$, $0<Q<1$, and $R$, $0<R<1$, be fixed parameters.
Define the   ensemble $\{N,t,Q\}$ of binary $( N\times t)$-matrices $X=(\x(1),\x(2),\dots \x(t))$,
where  columns $\x(i)$, $i\in[t]$, $t\eq\lfloor2^{ RN}\rfloor$,  
are chosen independently and equiprobably from the set consisting of ${N \choose\lfloor QN\rfloor}$ columns of the fixed weight $\lfloor QN\rfloor$. Fix two subsets
$\S,\L\subset[t]$ such that $|\S|=s$, $|\L|=\l$ and $\S\cap\L=\emptyset$.
From~\req{s_L1} it follows that for $\{N,t,Q\}$, the expectation
$\overline{|\B(s,\ell,X)|}$ of the number  $|\B(s,\ell,X)|$ is
$$
\overline{|\B(s,\ell,X)|}\,=\,|\P_s(t)|\,\Pr\left\{\S\in\B(s,\l,X)\right\}.
$$
Therefore, the expectation of the error probability for almost cover-free $(s,\ell)$-codes is
\beq{B_L}
{\cal E}^{(N)}(s,\ell,R,Q)\eq\,{|\P_s(t)|}^{-1}\,\overline{|\B(s,\ell,X)|}=\,
\Pr\left\{\S\in\B(s,\l,X)\right\},
\eeq
where the code size $t=\lfloor2^{RN}\rfloor$.
The evident {\em random coding upper bound} on the error probability~\req{e} for cover-free $(s,\ell)$-codes
is formulated as the following inequality:
\beq{eQ}
\epsilon(s,\ell,R,N)\eq\min\limits_{X\,:\,t=\lfloor2^{RN}\rfloor}\,\left\{\frac{|\B(s,\ell,X)|}{|\P_s(t)|}\right\}
\,\le\,{\cal E}^{(N)}(s,\ell,R,Q)\quad \text{for any }0<Q<1.
\eeq

The expectation
${\cal E}^{(N)}(s,\ell,R,Q)$ defined by~\req{B_L} can be represented as follows
\beq{B_L3}
{\cal E}^{(N)}(s,\ell,R,Q)\,\,\,=\sum\limits_{k=\lfloor QN\rfloor}^{\min\{N,\,s\lfloor QN\rfloor\}}
\Pr\left\{\S\in\B(s,\l,X)\left/\,\left|\bigvee_{i\in \S}\x(i)\right|=k\,\right.\right\}\cdot\P_2^{(N)}(s,Q,k)\le
$$
$$
\le \sum\limits_{k=\lfloor QN\rfloor}^{\min\{N,\,s\lfloor QN\rfloor\}}\,\P_2^{(N)}(s,Q,k) \cdot\min\left\{1;\, {t-s \choose \l}\P_1^{(N)}(\l,Q,k)\right\},
\eeq
where we  apply the total probability formula  and  the standard union bound for the conditional probability
$$
\Pr\left\{\bigcup\limits_i\,C_i\,\left/C\right.\right\}\,\le\,\min\left\{1\,;\,\sum\limits_i\Pr\{C_i/C\}\right\},
$$
and introduce the notations
\beq{P1s}
\P_1^{(N)}(\l,Q,k)\,\eq\,\Pr\left\{\bigvee\limits_{i\in\S}\x(i)\succeq \bigwedge\limits_{j\in\L}\x(j) \,\left/\,\left|\bigvee_{i\in \S}\x(i)\right|=k\,\right.\right\}
\eeq
and
\beq{k}
\P_2^{(N)}(s,Q,k)\,\eq\,\Pr\left\{\left|\bigvee_{i\in \S}\x(i)\right|=k\right\},\quad
\lfloor QN\rfloor\le k\le \min\{N, s\lfloor QN\rfloor\}.
\eeq
Let $k\eq\lfloor qN\rfloor$ and the functions
\beq{D}
\D(\l,Q, q)\eq\lim_{N\to\infty}\frac{-\log_2\left[\P_1^{(N)}(\l,Q,k) \right]}{N}
\eeq
and
\beq{A}
\A(s,Q,q)\eq\lim_{N\to\infty}\frac{-\log_2\left[\P_2^{(N)}(s,Q,k)\right]}{N}
\eeq
denote the exponents of the logarithmic asymptotic behavior for the probability of events~\req{P1s} and~\req{k} for the ensemble $\{N,t,Q\}$ respectively. Define $\hat q\eq1-(1-Q)^s$.

In Appendix we will prove 

\textbf{Lemma 1.}\quad {\em The function $\A(s, Q, q)$ of the parameter $q$, $Q<q<\min\{1, sQ\}$,
	defined by~$\req{A}$  can be represented in the  parametric
	form~	\beq{Ay}
	\A(s, Q, q) \eq (1-q) \log_2(1-q) + q \log_2 \[ \frac{Qy^s}{1-y} \] + sQ \log_2 \frac{1-y}{y} + sh(Q),
	\eeq
	\beq{ySmall}
	q=Q\frac{1-y^s}{1-y},\qquad 0<y<1.
	\eeq}
{\em In addition, the function $\A(s, Q, q)$ is $\cup$-convex, monotonically decreases  in the interval
	$(Q, 1 - (1 - Q)^s)$, monotonically increases  in the interval $(1 - (1 - Q)^s, \min\{1, sQ\})$
	and its unique minimal value which is equal to $0$ is attained at $q=\hat q\eq1-(1-Q)^s$, i.e.,}
$$
\min\limits_{Q<q<\min\{1, sQ\}}\,\A(s, Q, q)\,=\,\A(s, Q,\,\hat q)=0,\quad 0<Q<1.
$$

\textbf{Lemma 2.}\quad {\em For $\l\ge2$, the value of the function $\D(\l,Q,q)$ defined by \req{D} at point $q=\hat q$ is equal to
	$$
	\D(\l, Q, \hat q) =(1-Q)\,\l\,\log_2 z-(1-\hat q)\log_2[1-(1-z)^\l]+
	$$
	$$
	+\ell\left(\frac{(1-Q)}{z}(1-z)-\left(\frac{(1-Q)}{z}-\hat q\right)(1-z)^\l\right)\log_2[1-z]+\l h(Q),
	$$
	where $z$ is uniquely determined by the following equation
	$$
	Q=\frac{(1-z)(1-(1-z)^\ell)-(1-\hat q)z(1-z)^\ell}{1-(1-z)^\ell}.
	$$}

The inequality~\req{B_L3} and the random coding bound~\req{eQ}
imply that  the error probability exponent~\req{E} satisfies the inequality
\beq{ER}
\E(s,\ell,R)\,\ge\,\underline{\E}(s,\ell,R)\,\eq\,\max\limits_{0\le Q\le1}\,E(s,\ell,R,Q),
\eeq
\beq{EQ}
E(s,\ell,R,Q)\,\eq\,\min\limits_{Q < q < \min\{1, sQ\}}\;
\left\{\A(s,Q,q)+[\D(\l,Q,q)-\l\,R]^+\right\}.
\eeq
Lemma 1 states that  $\A(s,Q,q)>0$ if $q\neq\hat q$. In particular, the condition $q\neq\hat q$ implies $E(s,\ell,R,Q)>0$. Therefore, if $\l\,R<\D(\l,Q,\hat q)$ then $E(s,\ell,R,Q)>0$, what, in turn, means (see \req{Csld} and \req{ER}) that
$$
C(s,\ell)\ge \underline{C}(s,\ell)\eq\frac{1}{\l}\max\limits_{0\le Q\le 1}\D(\l,Q,\hat q),\quad  \text{where }\hat q=1-(1-Q)^s.
$$
Thus, the lower bound~\req{CSL} is established. 
$\qquad \square$

\textbf{Proof of Claim 2.}\quad
Let $\l\ge2$ be fixed and $s\to\infty$. Substituting $z=s/(s+\l)$ in \req{CSL}-\req{QZL} yields
$$
Q=\frac{(1-z)(1-(1-z)^\ell)-(1-\hat q)z(1-z)^\ell}{1-(1-z)^\ell}=\frac{\l}{s+\l}-\frac{\l^\l e^{-\l}}{s^\l}+O\left(\frac{1}{s^{\l+1}}\right),
$$
$$
\hat q=1-(1-Q)^s=1-e^{-\frac{s\l}{s+\l}+O\left(\frac{1}{s}\right)}=1-e^{-\l}+O\left(\frac{1}{s}\right)
$$
and
$$ 
C(s,\ell)\ge \frac{1}{\l}\max\limits_{0\le Q\le 1}\D(\l,Q,\hat q)=\frac{1}{\l}\max\limits_{0\le z\le 1}\D(\l,Q(z),\hat q(z))\ge\frac{1}{\l}\D(\l,Q(s/(s+\l)),\hat q(s/(s+\l))),
$$
where
$$
\D(\l, Q, \hat q) \eq (1-Q)\,\l\,\log_2 z-(1-\hat q)\log_2[1-(1-z)^\l]+
$$
$$
+\ell\left(\frac{(1-Q)}{z}(1-z)-\left(\frac{(1-Q)}{z}-\hat q\right)(1-z)^\l\right)\log_2[1-z]+\l h(Q).
$$
Therefore, one can write
$$
C(s,\ell)\ge \left(\frac{s}{s+\l}+\frac{\l^\l e^{-\l}}{s^\l}+O\left(\frac{1}{s^{\l+1}}\)\)\,\log_2\left[\frac{s}{s+\l}\right]-\(\frac{e^{-\l}}{\l}+O\(\frac{1}{s}\)\)\log_2\left[1-\left(\frac{\l}{s+\l}\right)^\l\right]+
$$
$$
+\(1+O\(\frac{1}{s^\l}\)\)\frac{\l}{s+\l}\log_2\left[\frac{\l}{s+\l}\right]-\({e^{-\l}}+O\(\frac{1}{s}\)\)\left(\frac{\l}{s+\l}\right)^\l\log_2\left[\frac{\l}{s+\l}\right]-
$$
$$
-\(\frac{\l}{s+\l}-\frac{\l^\l e^{-\l}}{s^\l}+O\left(\frac{1}{s^{\l+1}}\right)\)\log_2\[\frac{\l}{s+\l}-\frac{\l^\l e^{-\l}}{s^\l}+O\left(\frac{1}{s^{\l+1}}\right)\]-
$$
$$
-\(\frac{s}{s+\l}+\frac{\l^\l e^{-\l}}{s^\l}+O\left(\frac{1}{s^{\l+1}}\right)\)\log_2\[\frac{s}{s+\l}+\frac{\l^\l e^{-\l}}{s^\l}+O\left(\frac{1}{s^{\l+1}}\right)\]=
$$
$$
=\frac{\l^{\l-1}\log_2 e}{e^{\l}\,s^\l}+O\left(\frac{\log_2 s}{s^{\l+1}}\right).
$$
This completes the proof of Claim 2.$\qquad \square$\\
\newpage
\section{Appendix}

\textbf{Proof of Lemma 1.}\quad Let $s \ge 2$, $0 < Q < 1$, $Q < q < \min \{ 1, sQ \}$ be fixed parameters.
Assume also $k \eq \lfloor qN \rfloor$ and $N \to \infty$.
With the help of  the {\em type} (see~\cite{ck85}, \cite{d14_pit}) terminology:
$$
\{n(\a)\},\quad \a\eq(a_1,a_2,\dots,a_s)\in\{0,1\}^s,\quad 0\le n(\a)\le N,\quad
\sum\limits_{\a}n(\a)=N,
$$
the probability of event~\req{k} in the ensemble $\{N,t,Q\}$
can be written as follows:
\beq{k1}
\P_2^{(N)}(s,Q,k)\,= \,{N \choose \lfloor QN\rfloor }^{-s}\cdot  \,
\sum\limits_{\req{Qk}}\frac{N!}{\prod_{\a}n(\a)!},\quad
\lfloor QN\rfloor\le k\le \min\{N, s\lfloor QN\rfloor\},
\eeq
and in the right-hand side of~\req{k1}, the sum is taken over all types $\{n(\a)\}$ 
provided that
\beq{Qk}
n(\0)=N-k,\qquad \sum\limits_{\a:\,a_i=1}n(\a)=\lfloor QN\rfloor \quad \text{for any }i\in[s].
\eeq
For every type $\{n(\a)\}$ we will consider the corresponding distribution $\tau: \tau(\a) = \frac{n(\a)}{N}, \quad \forall~\a \in \{0, 1\}^s$. Applying the Stirling approximation, we obtain the following logarithmic asymptotic behavior of a term in the sum \req{k1}:
$$
-\log_2 \frac{N!}{\prod_{\a}n(\a)!} {N \choose \lfloor QN \rfloor}^{-s} = N F(\tau, Q, q)(1 + o(1)), \quad \text{where}
$$
\beq{FNeedToMin}
F(\tau, Q, q) = \sum_{\a}\tau(\a) \log_2 \tau(\a) + sH(Q).
\eeq
Thus, one can reduce the calculation of $\A(s, Q, q)$ defined by \req{A} to the search of the minimum:
\beq{FProblem}
\A(s, Q, q) = \min_{\tau \in \req{FRegion}: \req{FRestrictions}} F(\tau, Q, q)\eq F(\hat\tau, Q, q),
\eeq
\beq{FRegion}
\left\{ \tau:~\forall~\a \quad 0 < \tau(\a) < 1 \right\},
\eeq
\beq{FRestrictions}
\sum_{\a} \tau(\a) = 1, \qquad \tau(\0) = 1 - q, \qquad \sum_{\a: a_i = 1} \tau(\a) = Q \quad \forall~i \in [s],
\eeq
where the restrictions \req{FRestrictions} are induced by the definition of type and the properties \req{Qk}.

To find the minimum \req{FProblem} and the extremal distribution $\{\hat{\tau}\}$ we use  the method of Lagrange multipliers. The Lagrangian is
\begin{multline*}
\Lambda \eq \sum_{\tau(\a)} \tau(\a) \log_2 \tau(\a) + sh(Q) + \lambda_0 \(\tau(\0) + q - 1\) +\\
+ \sum_{i = 1}^s \lambda_i \(\sum_{\a: a_i = 1} \tau(\a) - Q\) + \lambda_{s+1} \(\sum_{\a} \tau(\a) - 1\).
\end{multline*}
Therefore, the necessary conditions for the extremal distribution $\{\hat{\tau}\}$ are
\beq{FNecConditions}
\begin{cases}
	\frac{\partial \Lambda}{\partial \tau(\0)} = \log_2 \hat\tau(\0) + \log_2 e + \lambda_0 + \lambda_{s+1} = 0,\\
	\frac{\partial \Lambda}{\partial \tau(\a)} = \log_2 \hat\tau(\a) + \log_2 e + \lambda_{s+1} + \sum_{i = 1}^s a_i \lambda_i = 0 \quad \text{for any } \a \neq \0.
\end{cases}
\eeq

It turns out that the matrix of second derivatives of the Lagrangian is diagonal and positive definite in the region \req{FRegion}, and the function $F(\tau, Q)$ defined by \req{FNeedToMin} is strictly $\cup$-convex in the region \req{FRegion}. The Karush-Kuhn-Tacker theorem  states that each solution $\tau \in \req{FRegion}$ satisfying system \req{FNecConditions} and constraints \req{FRestrictions} gives a local minimum of $F(\tau, Q)$. Thus, if there exists a solution of the system \req{FNecConditions} and \req{FRestrictions} in the region \req{FRegion}, then it is unique and gives a minimum in the minimization problem \req{FProblem} - \req{FRestrictions}.

Note that the symmetry of problem yields the equality $v \eq \lambda_1 = \lambda_2 = \dots = \lambda_s$. Let $u \eq \log_2 e + \lambda_{s+1}$ and $w \eq \lambda_0$. One can rewrite \req{FRestrictions} and \req{FNecConditions} as follows:
\beq{FFinalSystem}
\begin{cases}
	\text{1) } \log_2 \hat\tau(\a) + u + v \sum_{i = 1}^s a_i = 0 \quad \text{for any } \a \neq \0,\\
	\text{2) } \log_2 \hat\tau(\0) + u + w = 0,\\
	\text{3) } \hat\tau(\0) = 1 - q,\\
	\text{4) } \sum_{\a} \hat\tau(\a) = 1,\\
	\text{5) } \sum_{\a: a_i = 1} \hat\tau(\a) = Q \quad \text{for any } i \in [s].
\end{cases}
\eeq

Let $y \eq \frac{1}{1+2^{-v}}$.
The first equation of the system \req{FFinalSystem} means that
\beq{FTauANotLast}
 \hat\tau(\a) = \frac{1}{2^u y^s} (1-y)^{\sum a_j} y^{s - \sum a_j} \quad \text{for any } \a \neq \0 .
\eeq
Substituting \req{FTauANotLast} into the equation 5)  allows us to obtain
$$
\sum_{\a : a_i = 1} \frac{1}{2^u y^s} (1-y)^{\sum a_j} y^{s - \sum a_j} = \frac{1-y}{2^u y^s},
$$
and therefore the solution $u$ is determined by the equality
\beq{FVarU}
u = \log_2 \left[\frac{1 - y}{Qy^s}\right].
\eeq
Substituting \req{FTauANotLast}, \req{FVarU} and the third equation of \req{FFinalSystem} into the equation 4) of the system \req{FFinalSystem} we have
$$
q = \sum_{\a \neq 0} \hat\tau(\a) = \frac{Q(1 - y^s)}{1 - y},
$$
i.e. the equation \req{ySmall}. Thus, the conditions \req{FRestrictions} and \req{FNecConditions} have the unique solution $\tau$ in the region \req{FRegion}:
\beq{FTauSolution}
\hat\tau(\0) = 1 - q, \qquad \hat\tau(\a) = \frac{Q}{1 - y} (1-y)^{\sum a_j} y^{s - \sum a_j} \quad \text{for any } \a \neq \0,
\eeq
where the parameters $q$ and $y$ are related by the equation \req{ySmall}.
To get the exact formula \req{Ay}, the substitution of \req{FTauSolution} into \req{FNeedToMin} is sufficient.


Let us prove the properties of the function \req{Ay}.
Note that the function $q(y) = Q \frac{1 - y^s}{1 - y}$ \req{ySmall} monotonically increases in the interval $y \in (0, 1)$ and correspondingly takes the values $Q$ and $sQ$ at the ends of the interval. That is why one can consider the function \req{Ay} as the function $\F(s, Q, y) \eq \A(s, Q, q(y))$ of the parameter $y$ in the interval $y \in (0, y_1)$, where $q(y_1) = \min\{1, sQ\}$. The derivative of the function $\F(s, Q, y)$ equals
\beq{dLem1}
\F^{\prime}(s, Q, y) = q^{\prime}(y) \log_2 \left[\frac{Qy^s}{1 - Q - y + Qy^s}\right].
\eeq
Thus, $\F(s, Q, y)$ decreases in the interval $y \in (0, 1 - Q)$,
increases in the interval $y \in (1 - Q, y_1)$, is $\cup$-convex,
attains the minimal value $0$ at  $y_0 = 1 - Q$ and $q(y_0) = 1 - (1 - Q)^s$.$\qquad \square$

\textbf{Proof of Lemma 2.}\quad
Now, compute the conditional probability
$$
\P_1^{(N)}(\l,Q,k)\,\eq\,\Pr\left\{\bigvee\limits_{i\in\S}\x(i)\succeq \bigwedge\limits_{j\in\L}\x(j) \,\left/\,\left|\bigvee_{i\in \S}\x(i)\right|=k\,\right.\right\}
$$
Let $q,\,Q\le q\le\min\{1,sQ\}$, be fixed and $k\eq\lfloor qN \rfloor,\,\lfloor QN\rfloor\le k \le s\lfloor QN\rfloor$. In terms of \textit{types} (see~\cite{ck85}, \cite{d14_pit}):
\beq{type}
\{n(\a)\},\quad \a\eq(a_1,a_2,\dots,a_s)\in\{0,1\}^\l,\quad 0\le n(\a)\le N,\quad
\sum\limits_{\a\in\{0,1\}^\l}n(\a)=N,
\eeq
one can rewrite the probability in the following form
\beq{Pro}
\P_1^{(N)}(\l,Q,k)=\sum\limits_{\req{limPro}}\frac{N!}{\prod\limits_{\a\in\{0,1\}^\l}n(\a)!}\frac{{k\choose n(\1)}}{{N \choose n(\1)}}{N \choose \lfloor QN\rfloor }^{-\ell},\quad
\eeq
where the summation is taken over all choices of types $\{n(\a)\}$ provided that
\beq{limPro}
\sum_{\a:\,a_i=1}n(\a)=\lfloor QN\rfloor\quad\text{for any }i\in[\l].
\eeq
Applying the Stirling formula calculate the logarithmic behaviour of a term in~\req{Pro}

$$
\log_2\left[\frac{N!}{\prod\limits_{\a\in\{0,1\}^\l}n(\a)!}\frac{{k\choose n(\1)}}{{N \choose n(\1)}}{N \choose \lfloor QN\rfloor }^{-\ell}\right]=2^{-N F(\tau,Q,q)(1+o(1))},
$$
where
\beq{Ftq}
F(\tau,Q,q)\eq\sum_{\a\in\{0,1\}^\ell}\tau(\a)\log_2\tau(\a)-q\cdot h\left(\frac{\tau(\1)}{q}\right)+h(\tau(\1))+\ell\cdot h(Q).
\eeq
Here the \textit{probability distribution} $\{\tau(\a)\}$ is determined as
$$
\tau(\a)\eq\frac{n(\a)}{N}\quad\text{for any }\a\in\{0,1\}^\l.
$$
Since we are interested in 
$$\D(\l,Q,q)=\lim\limits_{N\to\infty}-\frac{\log_2 \left[P_1^{(N)}(\l,Q,k)\right]}{N},
$$
we might estimate the following minimum
\beq{FProblem}
\D(\l, Q, q) = \min_{\tau \in \req{FRegion}: \req{FRestrictions}} F(\tau, Q, q)\eq F(\hat\tau, Q, q),
\eeq
\beq{FRegion}
\left\{ \tau:~\forall~\a=(a_1,\dots, a_\l)\in\{0,1\}^\l \quad 0 < \tau(\a) < 1 \right\},
\eeq
\beq{FRestrictions}
\sum_{\a} \tau(\a) = 1,  \qquad \sum_{\a:\, a_i = 1} \tau(\a) = Q \quad \text{for any }i \in [\l],
\eeq
where the restrictions~\req{FRestrictions} are induced by properties~\req{type} and \req{type}.

To find the minimum we apply the standard Lagrange method, i.e., consider the Lagrangian
\begin{multline}\label{Lagr}
\Lambda\eq\sum_{\a\in\{0,1\}^\ell}\tau(\a)\log_2\tau(\a)-q\cdot h\left(\frac{\tau(\1)}{q}\right)+h(\tau(\1))+\ell\cdot h(Q)+\\
+\mu_0\cdot\left(\sum_{\a} \tau(\a) - 1\right)+\sum_{i=1}^{\l}\mu_i\cdot\left(\sum_{\a: \,a_i = 1} \tau(\a) - Q\right).
\end{multline}
Therefore, the necessary conditions for the extremal distribution $\{\hat\tau\}$ are
\beq{Der}
\begin{cases}
	\frac{\partial \Lambda}{\partial \tau(\a)} = \log_2 \hat\tau(\a) + \log_2 e + \mu_0 + \sum\limits_{i:\,a_i=1}\mu_i = 0\quad\text{for any } \a \neq \1,\\
	\frac{\partial \Lambda}{\partial \tau(\1)} = \log_2 \hat\tau(\1) + \log_2 e +  \sum\limits_{i = 0}^\l \mu_i +\log_2\left[\frac{1-\hat\tau(\1)}{q-\hat\tau(\1)}\right] = 0.
\end{cases}
\eeq
The matrix of second derivatives of the Lagrangian is obvious to be diagonal.  Thus, this matrix is positive definite in the region~\req{FRegion} and the function $F(\tau,Q,q)$ defined by~\req{Ftq} is strictly
$\cup$-convex in the region~\req{Ftq}. The Karush-Kuhn-Tacker theorem states
that each solution $\{\hat\tau\}$ satisfying system \req{Der} and constraints \req{FRestrictions} gives a local minimum
of $F(\tau,Q,q)$. Thus, if there exists a solution of the system \req{Der} and \req{FRestrictions} in the region~\req{FRegion}, then
it is unique and gives a minimum in the minimization problem~\req{FProblem}-\req{FRestrictions}.

Note that the symmetry of problem yields the equality $\mu \eq \mu_1 = \mu_2 = \dots= \mu_\l$. Let $\hat{\mu} \eq \log_2 e + \mu_{0}$. One can rewrite~\req{Der} as
\beq{sys}
\begin{cases}
	\hat{\mu }+\mu\sum_{i=1}^{\ell}a_i+\log_2[\hat\tau(\a)]=0\quad \text{for }\a\neq\1; \quad\cr
	\hat{\mu }+\mu\ell+\log_2[\hat\tau(\1)]+\log_2\left[\frac{1-\hat\tau(\1)}{q-\hat\tau(\1)}\right]=0;\cr
\end{cases}
\eeq
The first equations of~\req{sys} lead to
$$
\hat\tau(\a)=\frac{2^{-\hat{\mu}}}{z^\ell}\prod P(a_i) \quad \text{for } \a\neq\1,
$$
where we introduce the Bernoulli distribution
$$
P(a)\eq\begin{cases}
z\eq\frac{1}{1+2^{-{\mu}}}\quad \text{for }a=0;\cr
1-z\eq\frac{2^{-{\mu}}}{1+2^{-{\mu}}}\quad \text{for }a=1;\cr
\end{cases}
$$
In particular, it follows
\beq{mu}
\mu=\log_2\left[\frac{z}{1-z}\right].
\eeq
Since~\req{FRestrictions} the sum of all probabilities equals $1$ we get
\beq{t1}
\hat\tau(\1)=1-\sum_{k=0}^{\ell-1}{\l \choose k} \,\frac{2^{-\hat{\mu}}}{z^\ell}z^{\ell-k}(1-z)^{k}=1-\frac{2^{-\hat{\mu}}}{z^\ell}\left(1-(1-z)^{\ell}\right).
\eeq
The relation~\req{FRestrictions} of constant weight leads to
$$
Q=\frac{2^{-\hat{\mu}}}{z^\ell}\sum_{k=0}^{\ell-2}{\ell-1\choose k}z^{\ell-k-1}(1-z)^{k+1}+1-\frac{2^{-\hat{\mu}}}{z^\ell}\left(1-(1-z)^{\ell}\right)=1-\frac{2^{-\hat{\mu}}}{z^{\ell-1}}.
$$
This gives the connection between $\hat{\mu}$ and parameters $Q$ and $z$
\beq{mus}
\hat{\mu}=-\log_2\left[(1-Q)z^{\ell-1}\right].
\eeq
Finally, substituting~\req{mu}-\req{mus} to the second equation of~\req{sys} yields
\begin{multline*}
-\log_2\left[(1-Q)z^{\ell-1}\right]+\ell\log_2\left[\frac{z}{1-z}\right]+\log_2\left[1-\frac{(1-Q)}{z}\left(1-(1-z)^{\ell}\right)\right]+\\
+\log_2\left[\frac{(1-Q)}{z}\left(1-(1-z)^{\ell}\right)\right]-\log\left[q+\frac{(1-Q)}{z}\left(1-(1-z)^{\ell}\right)-1\right]=0
\end{multline*}
It can be written in the equivalent form
$$
\log_2\left[\frac{\left(1-(1-z)^{\ell}\right)}{(1-z)^\ell}\right]+\log_2\left[\frac{z-(1-Q)\left(1-(1-z)^\ell\right)}{(q-1)z+(1-Q)\left(1-(1-z)^\ell\right)}\right]=0
$$
This equation determines $Q$ as a function of parameters $z, \,q,\,s$ and $\ell$
\beq{Qus}
Q=\frac{(1-z)(1-(1-z)^\ell)-(1-q)z(1-z)^\ell}{1-(1-z)^\ell}.
\eeq
Notice that for fixed parameters $q,\,s$ and $\l $ there is a bijection between $Q\in[0,1]$ and $z\in[0,1]$.
From~\req{mus} and \req{Qus} it follows that
\beq{auxz}
\frac{2^{-\hat{\mu}}}{z^\ell}=\frac{1-Q}{z}=\frac{1-q(1-z)^\l}{1-(1-z)^\l}.
\eeq
Let us substitute $q=\hat q=1-(1-Q)^s.$ Thus
\beq{t1}
\hat\tau(\1)=\hat q(1-z)^\ell.
\eeq
Remind~\req{FProblem} that
\beq{FFF}
F(\hat{\tau},Q,\hat q)=\sum_{\a\in\{0,1\}^\ell}\hat\tau(\a)\log_2\hat\tau(\a)-\hat q\cdot h\left(\frac{\hat\tau(\1)}{\hat q}\right)+h(\hat\tau(\1))+\ell\cdot h(Q) .
\eeq
Let us rewrite the first sum of~\req{FFF} applying~\req{auxz}:
$$
\sum_{\a\in\{0,1\}^\ell}\hat\tau(\a)\log_2\hat\tau(\a)=\sum_{i=0}^{\l-1}{\ell \choose i}\frac{2^{-\hat{\mu}}}{z^\ell}(1-z)^i z^{\ell-i}\log_2\left[\frac{2^{-\hat{\mu}}}{z^\ell}(1-z)^i z^{\ell-i}\right]+\hat\tau(\1)\log_2\hat\tau(\1)=
$$
$$
=\,\sum_{i=0}^{\l-1}{\ell \choose i}\frac{2^{-\hat{\mu}}}{z^\ell}(1-z)^i z^{\ell-i}\log_2\left[\frac{2^{-\hat{\mu}}}{z^\ell}\right]+\sum_{i=0}^{\l-1}{\ell \choose i}\frac{2^{-\hat{\mu}}}{z^\ell}(1-z)^i z^{\ell-i}\log_2\left[z^{\ell-i}\right]+
$$
$$
+\sum_{i=0}^{\l-1}{\ell \choose i}\frac{2^{-\hat{\mu}}}{z^\ell}(1-z)^i z^{\ell-i}\log_2\left[{(1-z)^i}\right]+\hat\tau(\1)\log_2\hat\tau(\1)=
$$
$$
=\left(1-\hat  q(1-z)^\l\right)\log_2\left[\frac{1-\hat q(1-z)^\l}{1-(1-z)^\l}\right]+(1-Q)\,\l\,\log_2 z+
$$
$$
+\frac{(1-Q)}{z}\ell\left((1-z)-(1-z)^\l\right)\log_2[1-z]+\hat\tau(\1)\log_2\hat\tau(\1).
$$
Taking into account \req{t1} the second term of~\req{FFF} is
$$
-\hat q h\left(\frac{\hat\tau(\1)}{\hat q}\right)=\tau(\1)\log_2\left[\frac{\hat\tau(\1)}{q}\right]+(q-\hat\tau(\1))\log_2\left[\frac{q-\hat\tau(\1)}{q}\right]=
$$
$$
=\ell \hat q (1-z)^\l\log_2[1-z]+\hat q(1-(1-z)^\l)\log_2[1-(1-z)^\l].
$$
The third term of~\req{FFF} is 
$$
h(\hat\tau(\1))=-\hat\tau(\1)\log_2\hat\tau(\1)-(1-\hat\tau(\1))\log_2[1-\hat\tau(\1)].
$$
Finally, the last term of~\req{FFF} is $\l h(Q)$. Therefore, the value $\D(\l, Q, \hat q)= F(\hat\tau, Q, \hat q)$ can be written as
$$
\D(\l, Q, \hat q) = (1-Q)\,\l\,\log_2 z+\ell\left(\frac{(1-Q)}{z}(1-z)-\left(\frac{(1-Q)}{z}-\hat q\right)(1-z)^\l\right)\log_2[1-z]-
$$
$$
-(1-\hat q)\log_2[1-(1-z)^\l]+\l h(Q).
$$
Thus, we complete the proof of Lemma 2.$\qquad \square$

\end{document}